\documentclass{ws-ijmpb}

\begin{document}

\markboth{ Tanay Nag, Amit Dutta, and Ayoti Patra}
{quenching dynamics and quantum information }

%
\catchline{}{}{}{}{}
%

\title{QUENCHING DYNAMICS AND QUANTUM INFORMATION}

\author{TANAY NAG}
\address{Department of Physics, Indian Institute of Technology Kanpur, \\
Kanpur 208016,
India\\
tanayn@iitk.ac.in }

\author
{AMIT DUTTA}\address{Department of Physics, Indian Institute of Technology Kanpur, \\
Kanpur 208016,
India\\
dutta@iitk.ac.in }

\author{AYOTI PATRA}

\address{Department of Physics, University of Maryland,\\
 College Park,
Maryland 20742-4111, USA\\
ayoti@umd.edu }

\maketitle

\begin{history}
\received{Day Month Year}
\revised{Day Month Year}
\end{history}

\begin{abstract}
We review recent studies on the measures of zero temperature quantum correlations namely, the quantum entanglement (concurrence)  and discord present 
in the final state of a transverse $XY$ spin chain following a quench through quantum critical points; 
the aim of these studies is  to explore the scaling of the above quantities as a function of the quench rate.  
 A comparative study between the concurrence and  the quantum discord shows that their behavior is qualitatively 
the same though there are quantitative differences. For the present model, the scaling of both the
quantities are given by the scaling of the density of the defect present in the final state though one
can not find a closed form expression for the discord. We also extend our study of quantum discord to a transverse Ising chain in the presence of 
a three spin interaction. Finally, we present a study of the dynamical evolution of quantum discord and concurrence when two central qubits, initially prepared in a Werner state, are coupled to the environmental $XY$ spin chain which is  driven through quantum critical points. The qualitative behavior of quantum discord and concurrence are found to
be   similar as that of the decoherence factor.


\end{abstract}

\keywords{quantum phase transitions; non-equilibrium dynamics; concurrence; discord.}

\section{Introduction}

Quantum phase transition (QPT) is a zero-temperature transition of a quantum many body system
driven by quantum fluctuations arising due to the presence of non-commuting terms in the Hamiltonian \cite{sachdev99,chakrabarti96,continentino,vojta03}. A QPT is generally associated with a diverging length scale (correlation length, $\xi$) and a diverging time scale (relaxation time, $\xi_\tau$).  Both the $\xi$ and $\xi_\tau$ diverge in a power law fashion  as the quantum critical point (QCP), at $\lambda=0$, is approached by varying a parameter $\lambda$ of the Hamiltonian; $\xi \sim\lambda^{-\nu}$ and $\xi_\tau \sim \xi^{z}$. The notion of universality demands that  the correlation length exponent $\nu$ and the dynamical exponent $z$ depend on the dimensionality of the system, symmetry of the order parameter and the nature of the critical fixed point. In recent years, QPTs have been observed experimentally in a large number of systems, for example, in optical lattices  a Mott insulator to superfluid transition is observed \cite{makhlin00,greiner02,jaksch98}.

Recently, there has been a plethora of studies directed to understanding the non-equilibrium dynamics of quantum critical systems \cite{polkovnikov11,dutta10}. These theoretical studies 
are inspired by recent experiments on dynamic of  ultracold atomic gases \cite{sadler06,bloch08}. A natural
question to ask that what happens when a quantum system is driven across a QCP by changing
some parameter of the quantum Hamiltonian at a particular rate. It has been argued that even if
the system is prepared in  the ground state initially, there will be excitations (defects) in the final
state reached following a quench across the QCP. This can be attributed to the existence of a diverging time
scale in the vicinity of the QCP where the system becomes infinitely sluggish and hence can not
evolve adiabatically however slow the variation of the parameter be!   If a parameter $\lambda$ is
varied as $\lambda=t/\tau$, where $\tau$ is the inverse rate of quenching, the defect density ($\tilde n$)
in the final state is expected to satisfy a universal scaling relation given by $\tilde n \sim \tau^{-\nu d/(\nu z+1)}$ where $d$ is the spatial dimension and $\nu,z$ are the quantum critical exponents as
defined above; this is known as the Kibble-Zurek (KZ) scaling \cite{zurek96,zurek05,polkovnikov05} 
that has been extensively studied for transverse XY spin chain in recent years\cite{damski05,levitov06,mukherjee07} and 
 has also been extended to different quenching schemes, e.g., quenching through a 
multicritical point \cite{divakaran09,mukherjee10} (for recent reviews, see \cite{polkovnikov11,dutta10,dziarmaga10}).

Over the past few years, the connection between between quantum informations \cite{nielsen00,vedral07} and QPTs has
been explored extensively \cite{osterloh02}. It is interesting that  quantum information theoretic measures can capture the ground state singularities associated with a QPT. The quantum correlations of a state can be quantified in terms of bipartite entanglements. While the entanglement is a measure of the correlation based on the separability of two subsystems of a composite system, the quantum discord is based on
the measurement on one of the subsystems. Both the concurrence (which is one of the measures of bipartite entanglement) \cite{peres96,hill97,wootters98} and quantum discord \cite{olliver01} show distinctive behaviors close to the QCP of a one-dimensional transverse XY spin model and also interesting scaling relations which incorporate the information about the universality of the associated QPT. For a spin chain in the vicinity of a QCP, behavior and scaling of both concurrence
\cite{osterloh02} and discord \cite{dillen08,luo08,sarandy09,pal11}; for example, a second order derivative of quantum discord with respect to the driven parameter of the quantum Hamiltonian shows a peak at the QCP. We note that the behavior of entanglement entropy and Renyi entropy have been studied for transverse $XY$ spin chain for both equilibrium \cite{Franchini07,Franchini06,Franchini08,Franchini10} and in the non-equilibrium state following a quench \cite{Cincio07}.

Here, we provide an introductory review of the recent studies which in fact provide a bridge between the non-equilibrium dynamics of a quantum critical system and quantum information
theory. One may, for example, raise the question what is the value of concurrence or discord
in the final state of a quantum system following a quench across a QCP. If the dynamics
is perfectly adiabatic, then no additional correlation is generated in the final state. However,
as discussed above the  passage through a QCP invariably leads to defects in the final state and these 
defects in turn lead to non-zero entanglement \cite{sengupta09} and discord \cite{nag11}, i.e.,
non-adiabatic dynamics in the vicinity of the QCP results in  non-zero quantum correlations.
More importantly, both of them are found to scale with $\tau$ in an identical fashion to
that of the defect density.  Finally, we show that equilibrium and non-equilibrium dynamics of a spin chain can also alter the quantum correlation between two qubits initially in a Werner state \cite{werner89} which are
globally coupled to the chain \cite{yuan2007,liu2010}. These observations open up a possibility of a deep underlying connection between non-equilibrium quantum critical  dynamics and quantum information
theory.

The review is organized in the following way: in Sec. II we introduce the concept of quantum discord
and also briefly that of concurrence. In Sec. III., we introduce the phase diagram of the transverse
$XY$ spin chain and transverse Ising spin chain with the three spin interaction and quenching
schemes that are used in this review. In Sec. IV, the results for the transverse $XY$ spin chain
are presented and the possible scaling relation of concurrence and discord as a function of the
quenching rate is analyzed;  similar results for the three spin interacting model is given in the
next section. In Sec. VI, we discuss the dynamical evolution of concurrence and discord between two qubits coupled to a transverse XY spin chain pointing out the similarity of their behavior 
with that of the decoherence factor. Concluding remarks are presented in Sec. 7.

\section{Quantum Discord and Concurrence}

The notion of quantum discord 
is based on the idea that a measurement perturbs the system and prepares the system in suitable eigenbasis of that particular measurement operator. Two classically equivalent expressions of mutual information become non-equivalent in the quantum case when one of them is associated with a local measurement.  The difference between these two classically equivalent measures of 
mutual information is therefore the true indicator of quantum correlations present between two subsystems. This difference is in fact called the quantum discord.
\begin{figure}[bt]
\centerline{\psfig{file=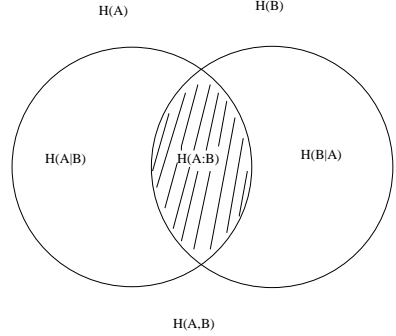,width=2.00in}}
\vspace*{8pt}
\caption{ Venn diagram explaining the notion of mutual information. 
$H(A)$ and $H(B)$ represent information associated with subsystems  $A$ and $B$ given in terms of the corresponding Shanon (von Neumann) entropies, respectively. $H(A,B)$ represents information associated with the composite system $AB$. $H(A|B)$ and $H(B|A)$ denote information associated with $A$ for a given information of $B$ and vice-versa, respectively. $H(A:B)$ signifies the mutual information associated with both $A$ and $B$.}
\label{Fig:venn}
\end{figure}

 To illustrate this concept, let us start with a classical bipartite system composed of two subsystems $A$ and $B$ (see Fig.~(\ref{Fig:venn})).  Let us try  to extract the  information about the composite system $AB$ from the overlapping part between $H(A)$ and  $H(B)$.
 The mutual information associated with the composite  system $AB$ is quantified in terms of Shannon entropy $H(p)$ where $p$ is the probability distribution of the combined system. The classical mutual information is defined as
  \begin{equation} 
  I(p)=H(p^A) + H(p^B) - H(p),
\label{total}
  \end{equation}
 where $H(p^i), i=A, B$ stand for the entropy associated with the subsystem  $i$ with probability
 $p^i$. Alternatively, this mutual information can be defined through the relation
   \begin{equation} 
  J(p)=H(p^A)-H(p|p^B),
  \end{equation}
  where $
  H(p|p^B)= H(p)-H(p^B) $
  is the conditional entropy.  In the classical situation, it is obvious that $I=J$ as any measurement on $B$ leaves the state of the composite system unchanged. 
  
Let us now attempt to generalize this to the quantum case when the classical Shannon entropy gets replaced by the quantum von Neumann entropy expressed in terms of the
density matrix. 
The natural quantum extension of Eq.~(\ref{total})  is
 \begin{equation}
 I(\rho)=s(\rho^A)+s(\rho^B)-s(\rho),
 \label{MI}
 \end{equation}
 where $\rho^A$ and $\rho^B$ are the density matrices for the subsystems $A$ and $B$, respectively, while  $\rho$ is that of the composite system.
In the quantum context, the conditional entropy is defined through a  local measurement carried over the  subsystem $B$. 
Usually one employs a von Neumann type of measurement consisting of a set of one dimensional projection operators $ \{\hat{ B_k }\}$. Following a local measurement on the subsystem $B$, the probability of the final
state $\rho_k$ of the composite system  where
 \begin{equation}
 \rho_k=\frac{1}{p_k}(\hat{I}\otimes \hat{B_k})\rho (\hat{I}\otimes \hat{B_k}),
  \end{equation}
 is  given by $p_k={\rm tr}(\hat{I}\otimes \hat{B_k})\rho(\hat{I}\otimes \hat{B_k})$, where $\hat{I}$ is the identity operator defined in the subspace of  $A$. One can therefore  define the quantum generalization of the conditional entropy given by $s(\rho|\{\hat{B_k}\})=\sum_k p_k s(\rho_k)$ leading to the equivalent definition of the  mutual information given by $J(\rho|\{\hat{B_k}\})=s(\rho^A)-s(\rho|\{ \hat{B_k}\})$.
 
 The classical correlation 
 can be extracted by maximizing the above expression  of expression of $J$ in the following way \cite{henderson01} 
  \begin{equation}
 C(\rho)=max _ {\{\hat{B_k}\}}  J(\rho| \{\hat{B_k}\}).
 \label{classical}
 \end{equation}
 As introduced by Olliver and Zurek \cite{olliver01}, the difference between  quantum mutual information $I(\rho)$ (Eq.(\ref{MI})) and the measurement induced classical correlation (Eq.(\ref{classical})), i.e.,
 \begin{equation}
 Q(\rho)= I(\rho)-C(\rho)
 \label{quantum}
 \end{equation}
is the quantum discord which is the true measures the quantum correlations.

It is worth noting that  $I$  represents the total information (correlation) whereas
$C$ is the information of  $A$ following a measurement in $B$. If $Q=0$, one can conclude  that the measurement has extracted all the information about the correlation between $A$ and $B$ i.e., system has no quantum correlation. On the other hand, a non-zero $Q$ implies  that entire information about the $A$
can not be extracted by local measurement on $B$. This difference originates from the fact that the subsystem $B$ gets disturbed in the process of measurement which does not happen
in classical information theory.

Let us quickly recap the notion of concurrence which is a measure of bipartite entanglement based on the separability approach. If states of the composite  system could be written in terms of tensor product of states of the subsystems $A$ and $B$ then the concurrence is zero i.e., they are unentangled. 
The concurrence for a mixed state of two qubits ($A$ and $B$) is given by $C_{nc} = {\rm max}
\{ 0, \sqrt{\lambda_1} - \sqrt{\lambda_2} - \sqrt{\lambda_3} -
\sqrt{\lambda_4} \}$, where $\lambda_i$'s are the eigenvalues of
$\rho (\sigma^y \otimes \sigma^y \rho^* \sigma^y \otimes \sigma^y) $ in
decreasing order \cite{peres96,hill97,wootters98} .  There could exist quantum states with zero entanglement but nonzero quantum discord.  As to be shown below, the final state of the spin chain following a quantum quench may represent this type of a state if the rate of quenching is below a threshold value.

\section{Model and Quenching Scheme}

\subsection{Transverse $XY$ spin chain}
\label{sec_txy}
The Hamiltonian of one-dimensional spin-1/2 $XY$ model in a transverse field with nearest neighbor ferromagnetic interactions is given by \cite{lieb61}
\begin{equation}
H = - \frac{1}{2} ~\sum_i ~[(1+\gamma) \sigma^i_x \sigma^{i+1}_x + (1- \gamma) \sigma^i_y \sigma^{i+1}_y + h \sigma^i_z],
\label{h1} 
\end{equation}
where $\sigma$'s are the Pauli spin matrices and the subscripts stand for the spin direction and superscripts are the lattice indices. 
The parameter $h$ is the transverse magnetic field  and $\gamma$ 
measures the anisotropic nature of the interactions; $\gamma=1$ corresponds to the transverse Ising chain\cite{chakrabarti96}.

\begin{figure}[bt]
\centerline{\psfig{file=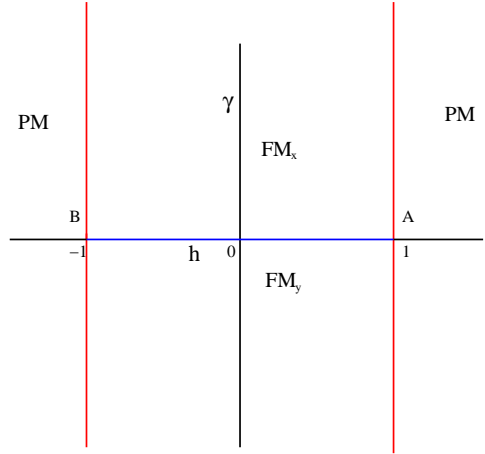,width=2.50in}}
\vspace*{8pt}
\caption{(Color online) The phase diagram of one dimensional $XY$ model in a transverse field given by Hamiltonian (\ref{h1}). The vertical red lines at $h= \pm 1$ denotes Ising transition from ferromagnetic phase to paramagnetic (PM) phase. The horizontal blue line stands for anisotropic phase transition between two ferromagnetic phases FM$_x$ and FM$_{y}$ with magnetic  ordering in $x$ and $y$ directions, respectively.
 }
\label{Fig:phase}
\end{figure}

The  model (\ref{h1}) can be exactly solved by mapping the spins to spinless non-interacting fermions via a Jordan-Wigner transformation \cite{lieb61} followed by a Bogoliubov transformation. The effective two-level Hamiltonian for each pair of
momenta $\pm k$ in terms of the states $|0 \rangle$ and
$|k,-k\rangle = c_k^\dagger c_{-k}^\dagger |0\rangle$ is given by 
\begin{equation}
H_k = 
\left(
\begin{array}{cc}
h+\cos k & \gamma \sin k\\
\gamma \sin k & -(h +\cos k)\\ 
\end{array}
\right),
\label{matform_h1}
\end{equation}
 where $c_k(c_k^{\dagger})$ is the Fourier transform of $c_i(c_i^{\dagger}) = \prod_{j=-\infty}^{i-1} \sigma_z^j
(-1)^i \sigma_-^{i} (\sigma_+^i)$; $|0 \rangle$, and $|k,-k\rangle$ denote the state with no and a pair of $c$-fermions, respectively. 
In the reduced Hilbert space, any general state can be represented as a 
superposition of $ |0\rangle$ and $|k,-k\rangle$ 
with time dependent amplitudes $u_k(t)$ and $v_k(t)$ such that
$\psi_k(t)=u_k(t)|0\rangle+v_k(t)|k,-k\rangle$. 
The phase diagram for the model is shown in Fig.~(\ref{Fig:phase}). The dynamical critical
exponents and the correlation length exponents associated with the Ising transitions (at $h=\pm 1$)
and the anisotropic transition ($\gamma =0$, $-1 \le h < +1$) are given by $\nu=z=1$.

We shall study the behavior of  quantum discord in the final state after quenching the system across  Ising critical points following the quenching scheme $h(t)=t/ \tau$, with
$t$ going from $-\infty$ to $\infty$ \cite{levitov06,mukherjee07}.  
The diverging relaxation time close to the QCPs at $h=\pm 1$ leads to defects in the final state. Initially  (at $t \to -\infty$), the system is at the ground state $| 0 \rangle$ where all the spins are aligned in the $-z$ direction.
 At $t\to \infty$, on the other hand,  the system is in an excited state due to non-adiabatic
 excitations in the vicinity  of the QCPs at $h=\pm 1$;  the probabilities of non adiabatic transition  for the mode $k$ is given by the Landau-Zener transition formula \cite{landau}
 \begin{equation} 
 p_k=|u_k(+\infty)|^2=\exp(-\pi \tau \gamma^2 \sin ^2 k).
 \label{pIsing}
 \end{equation}
In the limit $\tau \to \infty$, only the modes close to the critical modes ($k=0$ or  $k=\pi$) contribute to (\ref{pIsing}) and one arrives at a simplified form $p_k = \exp(-\pi \gamma^2 k^2 \tau)$. 

We further use a  quenching scheme that drives the system across a quantum multicritical point (MCP) ``A'' . For  a linear path,  $h$ and $\gamma$ are related as
\cite{mukherjee10}
\begin{equation}
h(\gamma) = 1+|\gamma(t)| sgn(t);  ~~~\gamma(t)=-\frac{t}{\tau},
\label{path}
\end{equation} 
with $t$ varying from $-\infty$ to $+\infty$, where the MCP is at $t=0$.   We investigate the scaling of discord in the final state following this quench
using the expression for the  probability of excitation $p_k$ which  in this case gets modified to
  \cite{divakaran09,mukherjee10}
  \begin{equation} 
  p_k=\exp(- \pi \tau (1+\cos k)^{2} \sin ^2 k).
\label{pMCP}
\end{equation}

\subsection{Transverse Ising model with a three spin interaction}
Let us now introduce  a one-dimensional three spin interacting transverse Ising 
system described by the Hamiltonian
\cite{kopp05,divakaran07}
\begin{equation}
H=-\frac 1 {2} \{\sum_{i}\sigma^{i}_{z}[h+J_3\sigma^{i-1}_{x}\sigma^{i+1}_{x}] -J_x\sum_{i}\sigma^{i}_{x}\sigma^{i+1}_{x}\},
\label{3spin}
\end{equation}
where $J_x$ is the strength of the nearest neighbor
ferromagnetic interaction and $J_3$ denotes the strength
of the three spin interaction. In the limit $J_3\to 0$, the model
reduces to transverse Ising model. Moreover,   Hamiltonian (\ref{3spin})
can be mapped to a transverse $XY$ spin chain with competing (ferroÐ antiferromagnetic) interactions in the $x$ and $y$ components of the spin  using  a duality transformation \cite{kopp05}. Therefore, it can be exactly solved using JW
transformations in spite of the presence of the three spin interactions.

\begin{figure}[bt]
\centerline{\psfig{file=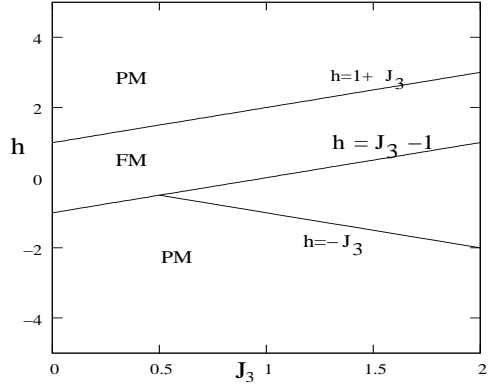,width=2.50in,height=2.0in}}
\vspace*{8pt}
\caption{Equilibrium phase diagram of the three spin interacting Ising model (\ref{3spin}).
Solid lines show phase boundaries between PM and FM phases.} 
\label{Fig:PH}
\end{figure}

The reduced Hamiltonian in the Fourier space
is of the form
\[\left[\begin{array}{ll}
(h(t)+J_x\cos k-J_3\cos 2k) &~~~~~~~ i (J_x\sin k-J_3\sin 2k)\\
-i(J_x\sin k-J_3\sin 2k)&-(h(t)+J_x\cos k-J_3\cos 2k)\end{array}\right]\]
We can analyze the gap of the spectrum and it can be seen that gap vanishes at $h=J_3+1$, and 
also at $h=J_3-1$,  with ordering 
 wave vectors $\pi$ and $0$, respectively; these transitions are equivalent to the Ising transitions
 in the transverse XY spin chain (\ref{h1}) while
the phase transition at $h=-J_3$ (and $J_3> 0.5$) belongs to the universality
class of the anisotropic transition. 
The equilibrium phase diagram is shown in the Fig.~(\ref{Fig:PH}) \cite{kopp05}. 

Let us consider the quenching dynamics of the system across  Ising critical points following the quench scheme $h(t)=t/ \tau$ with
$t$ going from $-\infty$ to $\infty$ \cite{divakaran07}.
The probability of non-adiabatic transition close to the quantum critical points are again given by  the corresponding Landau-Zener formula \cite{divakaran07}
\begin{equation}
p_k=\exp[-\pi\tau(\sin k-J_3\sin 2k)^2].
\end{equation}

\section{Pairwise Correlations, Quantum Discord, and Concurrence}

Let us now address the question what would be  concurrence and discord in the final paramagnetic phase of the Hamiltonians (\ref{h1}) following quantum quenches using the scheme $h(t)=t/\tau$.   The initial state ($t \to -\infty$)  is a direct product state with all the spins pointing in $-z$ direction. If the dynamics is perfectly adiabatic, the final state is also expected to be a direct product state (with
all spins pointing up). This implies that there is no quantum correlation generated if the dynamics
is perfectly adiabatic which is the never the case if the system crosses a QCP during the temporal
evolution. Passage through a QCP generates defects in the final state which in turn leads to
non-zero quantum correlations that show up in the measures like concurrence and discord.

To calculate these measures,  we use the  two-spin density matrix  for spins at the sites $i$ and $j=i+n$. Let us  recall  the generic form of the  density matrix given by 
\cite{sarandy09,syljuasen03}

\begin{equation}
 \rho^n = \frac{1}{4} (I^i \otimes I^j+c_1 \sigma_x^i \otimes \sigma_x^j + c_2 \sigma_y^i \otimes \sigma_y^j
+ c_3 \sigma_z^i \otimes \sigma_z^j 
 + c_4 I^i \otimes \sigma_z^j + c_5 \sigma_z^i \otimes I^j ),
\end{equation}
where 
   $c_1=\langle \sigma_x^i \sigma_x^{j} \rangle$, $c_2=\langle \sigma_y^i \sigma_y^{j} \rangle$, $c_3=\langle \sigma_z^i \sigma_z^{j} \rangle$,
 $c_4=c_5=\langle \sigma_z^i  \rangle$. For the Hamiltonian (\ref{h1}), the equivalence of $X$ and $Y$ directions (in the spin space) demands $c_1 = c_2$. The density matrix can also be  rewritten in the form
\begin{equation}
\rho^n = 
\left(
\begin{array}{cccc}
a^n_+ & 0 & 0 & b^n_1\\
0 & a^n_0 & b_2^n & 0\\  
0 & b^{n*}_2 & a_0^n & 0\\ 
b^{n*}_1 & 0 & 0 & a^n_-\\ 
\end{array}
\right),
\label{rdm}
\end{equation}
where the matrix elements are given in terms of the two-spin correlation functions in the following manner:
\begin{eqnarray}
a_{\pm}^n &=& \frac{1}{4} \langle (1 \pm \sigma^i_z)(1 \pm \sigma^{i+n}_z) \rangle = 1+c_3 \pm 2c_4, \nonumber \\
a_0^n &=& \frac{1}{4} \langle (1 \pm \sigma^i_z)(1 \mp \sigma^{i+n}_z)  \rangle =1-c_3, \nonumber \\
b_{1(2)}^n &=& \langle \sigma^i_- \sigma^{i+n}_{-(+)} \rangle. 
\label{elements}
\end{eqnarray}
Moreover, the up-down symmetry of the Hamiltonian leads to a simplified form of the density matrix
where some elements vanish \cite{syljuasen03}.
 
Defining a quantity  
\cite{levitov06,sengupta09}
\begin{equation}
\beta_n = \int_0^{\pi} \frac{dk}{\pi} p_k \cos(nk),
\label{alpha_n}
\end{equation}
one gets
\begin{eqnarray}
c_4 = c_5 = \langle \sigma_z^i \rangle &=& 1 -2 \beta_0 ,\nonumber \\
c_3 = \langle \sigma_z^i \sigma_z^{i+n} \rangle &=& \langle \sigma_z^i \rangle^2 - 4 \beta_n^2.
\label{dcorr}
\end{eqnarray}
The expressions for $c_1$ and $c_2$ can be computed  for different value of $n$ which we present below for $n \le 6$:
\begin{eqnarray}
c_1& =& c_2 = 
\left\{
\begin{array}{lr}
\frac{\beta_2}{2}(1-2 \beta_0), & n=2, \\
\\
(1-2 \beta_0)^2  \beta_2^2 - 4 \beta_2^4 + 
\frac{\beta_4}{2} (1-2 \beta_0)^3 - 2 \beta_2^2 \beta_4  (1-2 \beta_0),& n=4,\\
\\
\frac{1}{2}[ \beta_6 \{ (1 - 2 \beta_0)^2 - 4 \beta_2^2) \} + 4 \beta_2 \{ \beta_2^2 + 
  \beta_4^2 - \beta_4 (1 - 2 \beta_0)\}] \times [16 \beta_2^2 \beta_4 +\nonumber \\
 (1 - 2 \beta_0) \{ (1 - 2 \beta_0)^2 - 8 \beta_2^2 - 4 \beta_4^2\}], & n=6. \nonumber \\
\end{array} 
\right.
\end{eqnarray}
The eigen values of the density matrix are obtained in terms of the correlators $c_i$s as \cite{luo08,sarandy09} 
\begin{eqnarray}
  \lambda_0 &=& \frac{1}{4}[(1+c_3)+\sqrt{4c_4^2+(c_1-c_2)^2}], \nonumber \\
  \lambda_1 &=& \frac{1}{4}[(1+c_3)-\sqrt{4c_4^2+(c_1-c_2)^2}], \nonumber\\
  \lambda_2 &=& \frac{1}{4}[(1-c_3)+(c_1+c_2)], ~~\rm{and} \nonumber \\ 
  \lambda_3 &=& \frac{1}{4}[(1-c_3)-(c_1+c_2)],
  \label{eigenval}
  \end{eqnarray}
which can be expressed  in terms of $\beta$'s using the equations (\ref{elements}), (\ref{alpha_n}) and (\ref{dcorr}).
Using Eq.~(\ref{alpha_n}) we note that $\beta_n=0$  for odd $n$ as $p_k$ is invariant under $k \to \pi - k$. At the same time, 
$\langle \sigma_{\pm}^i \sigma_{\pm}^{i+n} \rangle= b_1^n=0$ for all $n$ since the expectation values  of a pair  of fermionic annihilation or creation operators do always vanish. Moreover, $\langle \sigma_{\pm}^i \sigma_{\mp}^{i+n} \rangle= b_2^n=0$ for odd $n$ since the quantities $b_2^n$ change sign under the $\mathbb{Z}_2$ 
transformation \cite{levitov06,sengupta09,syljuasen03}.

On the other hand, $b_2^n= c_1+c_2$ for even $n$. 
The variation of mutual Information $I$, the classical correlation $C$ and the 
quantum discord $Q = I-C$ with $\tau$ are therefore studied for both critical and multicritical quenches (\ref{path}) for even $n$.
Let us assume that the spin at site $i$ to be subsystem $A$ and at site $j$ as subsystem $B$. The reduced density matrix for the subsystems $A$ and $B$  can be expressed as
 \begin{eqnarray}
 \rho_A &=&  \frac{1}{2}( I^i \otimes I^j + c_4 I^i \otimes \sigma_z^j ), ~\rm{and} \nonumber \\
 \rho_B &=& \frac{1}{2}( I^i \otimes I^j + c_4  \sigma_z^i \otimes I^j  ),
 \end{eqnarray}
 with eigenvalues 
 \begin{eqnarray}
 \lambda_4 &=& \frac{1}{2}(1+c_4), ~\rm{and}\nonumber \\
 \lambda_5 &=& \frac{1}{2}(1-c_4).
 \end{eqnarray}
  The total mutual information  $I(\rho)$ is expressed terms of von Neumann entropies, which when substituted in Eq.~(\ref{total}) gives
 \begin{eqnarray}
 I(\rho)=s(\rho_A)+s(\rho_B)-\sum_{\alpha=0}^3 \lambda_\alpha \log_2 \lambda_\alpha , 
 \end{eqnarray}
 where
 $ 
 s(\rho_A)=s(\rho_B)=-\lambda_4 \log_2 \lambda_4 - \lambda_5 \log_2 \lambda_5 
 $. In order to calculate the classical correlation, we employ a set of projector for local measurement on the subsystem $B$ given by 
   $B_k=V\Pi_kV^\dagger$ where $\Pi_k= |k \rangle  \langle k| : k=+,-$ is the set of projectors on the computational basis
 $|+\rangle=\frac{1}{\sqrt{2}}( |0\rangle + |1\rangle), |-\rangle=\frac{1}{\sqrt{2}}( |0\rangle - |1\rangle)$ and $V \in U(2)$ where  $V$ is parametrized
over a  Bloch sphere given by
\begin{eqnarray}
\left(
\begin{array}{cc}
\cos \frac{\theta}{2}  &  \sin \frac{\theta}{2} e^{-i\phi}\\ 
\sin \frac{\theta}{2} e^{i\phi}  &  -\cos \frac{\theta}{2}\\ 
\end{array}
\right),
\end{eqnarray}
where the polar angle $\theta$ lies between  $0$ and  $\pi$ and the azimuthal angle $\phi$ can be varied from  $0$ to $ 2\pi$.
Following the standard techniques
\cite{sarandy09}
, we can obtain the classical correlation by maximizing
\begin{equation}
C(\rho)= s(\rho_A)-s(\rho_{+}),
\end{equation}
where $\rho_{+}$ is the density matrix for the outcome $ |k\rangle=|+\rangle$. Below we summarize the final results for $n=2$.
    
The exact expressions for mutual information and classical correlation are given below. 
\begin{eqnarray}
I &=& -2(1- \beta_0) \log_2(1 - \beta_0) 
    + ((1 - \beta_0)^2 - \beta_2^2)  \log_2((1- \beta_0)^2 - \beta_2^2) -2\beta_0 \log_2(\beta_0) \nonumber \\
  &+& (\beta_0^2 - \beta_2^2) \log_2(\beta_0^2 - \beta_2^2) 
    + \frac{1}{4}\{ 4 \beta_0(1 - \beta_0)
      + 4\beta_2^2 + \beta_2(1 - 2\beta_0)\}  \nonumber \\
      &\times & \log_2 \left[ \frac{1}{4}\{4 \beta_0(1 - \beta_0) + 4\beta_2^2 + \beta_2(1 - 2\beta_0)\} \right] 
     + \frac{1}{4}\{4 \beta_0(1 - \beta_0) + 4\beta_2^2 - \beta_2(1 - 2\beta_0)\} \nonumber \\
    & \times & \log_2\left[ \frac{1}{4} \{ 4 \beta_0(1 - \beta_0) + 4\beta_2^2 - \beta_2(1 - 2\beta_0) \} \right], \nonumber \\
\label{info2} 
\end{eqnarray}
and
\begin{eqnarray}
 C &=& -(1- \beta_0) \log_2(1 - \beta_0) - \beta_0 \log_2(\beta_0) 
     + \frac{1}{2} \left( 1 - (1 - 2\beta_0) \sqrt{ 1+ \frac{\beta_2^2}{4}} \right)  \nonumber \\
    & \times &\log_2 \left[ \frac{1}{2}\left \{ 1 - (1 - 2\beta_0) \sqrt{ 1+ \frac{\beta_2^2}{4}} \right \} \right]
  + \frac{1}{2} \left( 1 + (1 - 2\beta_0) \sqrt{ 1+ \frac{\beta_2^2}{4}} \right)  \nonumber \\
    & \times &\log_2 \left[ \frac{1}{2} \left \{1 + (1 - 2\beta_0) \sqrt{ 1+ \frac{\beta_2^2}{4}} \right \} \right].
\label{class2} 
\end{eqnarray}
Similarly one can obtain the expressions for $n=4$ and $n=6$ using the appropriate equations.
We note that  $I$ and $C$ and hence $Q=I-C$ depends entirely on  $\beta$'s which are in turn dependent on the quench rate $\tau^{-1}$ through the defect density $p_k$.

The spin chain has non zero concurrence \cite{sengupta09,syljuasen03} if $|b_2^n|>\sqrt{a_+^n a_-^n}$. The concurrence is given by $C_{nc}=\rm max \it \left \{0,2(|b_2^n|-\sqrt{a_+^n a_-^n})\right \}$.

\subsection{Comparative study between discord and concurrence}

\begin{figure}[bt]
\centerline{\psfig{file=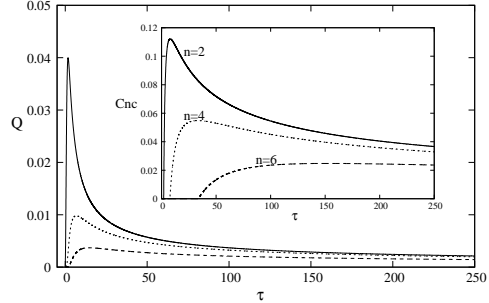,width=2.50in}}
\vspace*{8pt}
\caption{ The variation of quantum discord $Q$ with  $\tau$ for $n=2$ (solid line),$4$ (dotted line) and $6$(dashed line) in the final state following a linear quench across the Ising critical points where $\gamma=1$. Inset shows the variation of concurrence ($C_{\rm nc}$) for same parameter value.}
\label{Fig:QC}
\end{figure}

We would like to make a comparative study between the discord and concurrence as a function of quenching rate $\tau$. The variation of quantum discord $Q$ with $\tau$ for $n=2,4,6$  following a quench across the Ising critical point is shown in Fig.~(\ref{Fig:QC}).
  $Q$ vanishes in both the limits $\tau \to 0$ (sudden limit) and $\tau \to \infty$ (adiabatic limit) due to the fact that the final state is nearly a direct product state in either cases. In the adiabatic limit,
  the system nearly reaches the expected final state with minimum defect while in the sudden
  limit, the final state is almost identical to the initial ground state (which is a direct product state).
Therefore $Q$   increases monotonically with $\tau$ and reaches a peak value at an
intermediate $\tau = \tau^m$, and starts decreasing for $\tau >\tau^m$. A similar behavior is
also seen for  von-Neumann entropy density as shown in \cite{levitov06,mukherjee07}.

\begin{figure}[bt]
\centerline{\psfig{file=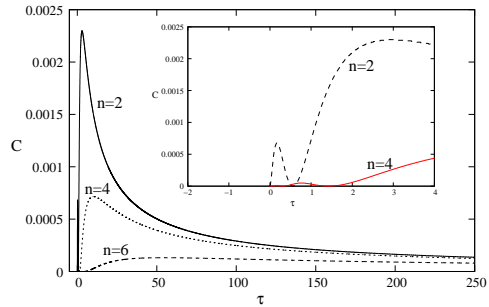,width=2.50in}}
\vspace*{8pt}
\caption{(color online)Variation of classical correlation with $\tau$ for $n=2$ (solid line),$4$ (dotted line) and $6$(dashed line). Inset shows  small peaks for $\tau \to 0$.}
\label{Fig:COsc}
\end{figure}

  As the lattice spacing $n$ increases, $\tau^m$ is also shifted to higher values and the magnitude
  of $Q$ decreases indicating that the correlation generated through quenching is short-ranged. 
  Classical correlation as defined in Eq.~(\ref{class2}) also exhibits a qualitatively identical behavior though it is smaller in magnitude in comparison to discord  (Fig.~(\ref{Fig:COsc})). It is worth noting, the classical correlations however show some initial fluctuating behavior for $\tau \to 0$  followed by the monotonic increase (see inset Fig.~(\ref{Fig:COsc})). 
This is due to the fact that the classical correlation becomes zero at some intermediate times when the measurement basis matches with the state of a subsystem, namely $i$-th spin. The value of $C$ is found to be one order of magnitude less than $Q$ implying that correlations present in the system are essentially quantum mechanical. 

The variation of concurrence in the final state for a quench across an Ising critical point with   $\tau$ has been studied recently 
\cite{sengupta09}
and the comparison is shown in Fig.~(\ref{Fig:QC}). Although the variation of discord and concurrence are qualitatively similar, we emphasize following differences.
The magnitude of discord is less than that of concurrence for the same value of  $n$ nearly by one order suggesting that discord represents quantum correlations.  Moreover, $Q$ shows a peak at a value of  $\tau^m$ which is very small in comparison to the 
corresponding $\tau^m$ for concurrence. Based on the above observations, we conclude that the measurement based approach employed in calculating discord provides a quantitatively different result for
quantum correlations. Moreover, the study on concurrence 
\cite{sengupta09} 
indicates the existence of  a threshold value of $\tau$  above which the  bipartite entanglement is generated through quenching.
On the contrary,  discord is non-zero for all $\tau$ as we observe negligible shift close to $\tau=0$ for different $n$ as shown in Fig.~(\ref{Fig:QC}).  Therefore, in some situations quenching leads to a final state that happens an ideal example of a quantum state with zero concurrence but non-zero discord.

\subsection{Scaling of discord and concurrence}

\begin{figure}[bt]
\centerline{\psfig{file=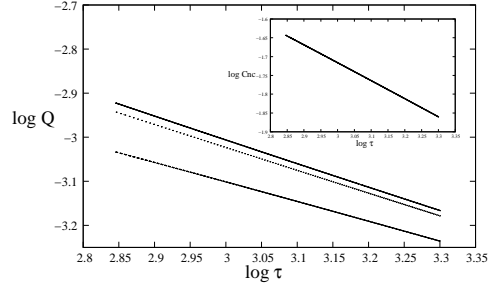,width=2.50in}}
\vspace*{8pt}
\caption{The Variation of  quantum discords $Q$ as a function of $\tau$ for $n=2$
(solid line),$4$ (dotted line) and $6$(dashed line) are shown on a
log-scale following a quench across the Ising critical points
with $\gamma=1$; the slope is $\approx 0.5$ in all cases indicating that discord satisfies identical scaling relation as that defect density. 
Inset shows the variation of concurrence ($C_{\rm nc}$)
for $n=2$ with the same parameter values
; the slope is again $\approx 0.5$}
\label{Fig:isingcomp}
\end{figure}

We would now like to explore whether $Q$  satisfies a universal scaling relation as a function of $\tau$; from the KZ scaling relation we recall that the defect density scales as $1/\sqrt {\tau}$ for quenching through the Ising critical points. It has been reported  that concurrence  also scales as 
   $1/{\sqrt \tau}$ in the limit of large $\tau$ \cite{sengupta09}. In Fig.~(\ref{Fig:isingcomp})we analyze  the scaling of discord which clearly shows a $1/{\sqrt \tau}$ fall.   
   In order to explore the scaling of $Q$ analytically in the limit $\tau \to \infty$,
we analyze the asymptotic behavior of the terms of  $I$  (Eq.~(\ref{info2})). 
The first two terms together decay as  $1/\tau$,  while the contribution from the remaining four terms has been found to scale numerically as $1/\sqrt {\tau}$, which determine the scaling when
 $\tau \to \infty$. Although, a closed power-law form is not obtained, our
studies apparently points to the fact that discord does also satisfy a scaling analogous to that of concurrence or defect density. 

\begin{figure}[bt]
\centerline{\psfig{file=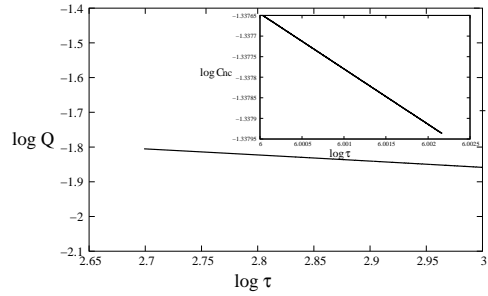,width=2.50in}}
\vspace*{8pt}
\caption{The variation of discord for $n=2$
following a multicritical quench along a linear path with
 slope 
 $\approx -0.19$  which matches well with the
value of the exponent $-1/6$ obtained for the defect density
. In the inset, we show the similar variation of
concurrence with $\tau$ for $n=2$ %
 and the slope is $\approx
-0.13$ which is in close agreement with the exponent $-1/6$.}
\label{Fig:multicomp}
\end{figure}

We further verify this claim by investigating the scaling of discord following a linear quenching across the MCP `A' (see Eq.~(9))
for which the defect density scales as $\tau^{-1/6}$.  
\cite{divakaran09,mukherjee10}  
 In Fig.~(\ref{Fig:multicomp}), we present  the scaling of discord (and also that of concurrence) with respect to $\tau$; our numerical results again indicates
that the scaling of discord is likely to be same as that of the defect density.

\section{Some Results for Transverse Ising spin chain with three Spin Interaction}

\begin{figure}[bt]
\centerline{\psfig{file=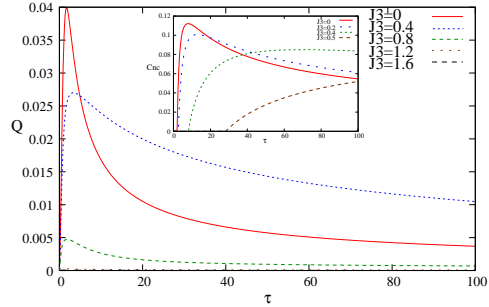,width=2.50in}}
\vspace*{8pt}
\caption{(color online)The Variation of  quantum discords $Q$ for $n=2$ as a function of $\tau$ with different values of $J_3$.
Discord decays slowly around $J_3=0.5$ than any other values of $J_3$. Inset shows the Variation of concurrence as a function of $\tau$ with different values of $J_3$ for the same lattice spacing $n=2$.
Concurrence becomes zero when $J_3>0.5$.}
\label{Fig:dis}
\end{figure}

\begin{figure}[bt]
\centerline{\psfig{file=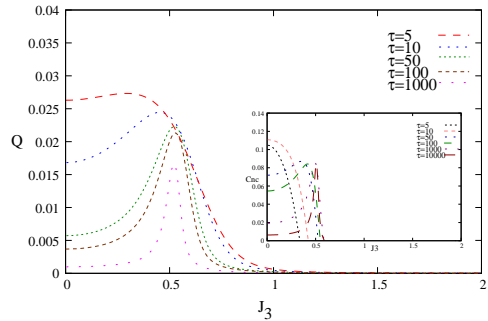,width=2.50in}}
\vspace*{8pt}
\caption{(color online)The Variation of $Q$ for $n=2$ as a function of $J_3$ with different values of $\tau$. 
 A sharp peak of discord gradually approaches around $J_3=0.5$ as $\tau$ increases. The inset shows the Variation of concurrence as a function of $J_3$ for the above parameter values. Concurrence shows a peak around $J_3=0.5$ and becomes zero as $J_3>0.5$ for any value of $\tau$.}
\label{Fig:vsj3}
\end{figure}

In this section, we extend our previous study (in Sec.IV.) to the context of a three spin interacting transverse Ising chain (\ref{3spin}) for quenching scheme $h(t)=t/\tau$.  Using  reduced
matrix (\ref{rdm}) for the present model, one obtains
the variation of quantum discord $Q$ as a function of $\tau$ with  different values of $J_3$
as shown in 
Fig.~(\ref{Fig:dis}). 
The peak hight of discord decreases with increasing three spin interaction strength.
This can be explained by noting that the magnitude of the defect density  is inversely related to $J_3$\cite{divakaran07} and discord, as discussed previously, is expected to be proportional to the
defect density.
The Inset of Fig.~(\ref{Fig:dis}) shows the variation of concurrence as a function of quench rate. Although the behavior of both of them are similar,  the  concurrence vanishes if $J_3$ exceeds $0.5$.
One can further check that the discord and concurrence satisfy the same scaling as the defect
density for both critical and multicritical quenches.

Fig.~(\ref{Fig:vsj3})  shows the variation of quantum discord and concurrence as a function of $J_3$ for a fixed value of quench time $\tau$.
 Discord shows a peak at a value of $J_3$ that approaches $J_3 =0.5$ and becomes sharper with increasing $\tau$;  on the other hand, concurrences vanishes for $J_3 >0.5$ for all values of $\tau$.

\section{Dynamical Evolution of Quantum Discord and Concurrence}
In the earlier section, we  presented a review of the scaling of concurrence and discord in the final state following a quench across a QCP and their
connection to the scaling of the defect density. In this section, on the other hand, we  study  the dynamical evolution of these measures 
between a pair of qubits which are coupled to a driven quantum spin chain; we choose the environmental spin chain to be the transverse XY
spin chain defined in Eq.~(\ref{h1}).
This environmental spin chain is globally coupled to the  two qubits (denoted by $A$ and $B$)  through  the  interaction Hamiltonian given by  $H_{SE}= \sum_i^N \delta/2 (S_z^{A} + S_z^{B})\sigma^i_z $ where $\delta$ 
denotes the coupling strength
between system and environment \cite{yuan2007,liu2010}. 
We note  that $[S_z^{A} + S_z^{B},\sigma_{\alpha}^{i}] = 0 ~(\alpha = x, y, z)$ so that $\delta/2 (S_z^{A} + S_z^{B})$ is  conserved during time evolution.
The global Hamiltonian of the system and qubits can then be written as,
\begin{equation}
H = \sum_{\mu=1}^4 |{\varphi_\mu}\rangle \langle{\varphi_\mu}| \otimes H_{E}^{h_\mu}(h(t)+ \varepsilon_{\mu}),
\label{product}
\end{equation}
with $|{\varphi_\mu}\rangle(\mu = 1,2,3,4; ~|{\varphi_1}\rangle= |++\rangle ,
 |{\varphi_2}\rangle= |--\rangle, |{\varphi_{3,4}}\rangle= 1/\sqrt{2}(|+-\rangle \pm |-+\rangle))$ 
denoting the $\mu$th eigenstate of the operator $ (S_z^{A} + S_z^{B})$, i.e., central two qubits system, with the $\mu$th eigenvalue $\varepsilon_\mu (\varepsilon_1 = +\delta, \varepsilon_2 =-\delta \rm, \varepsilon_{3,4}=0)$. The parameters 
$h_\mu$ are given by $h_\mu(t) = h(t) + \varepsilon_\mu$, and the   environmental Hamiltonian
 $H_{E}^{h_\mu}$ is obtained from $H_{E}$ by replacing 
$h(t)$ with $h_\mu(t)$. We therefore find that the coupling between the qubits and the environment, splits the Hamiltonian into two branches with
transverse field $(h(t)+\delta)$ and $(h(t)-\delta)$, respectively. For our studies, employ the following time variation of of the transverse field  $h(t)$ given by $h(t)=1-t/\tau$ while $t$ varies from 
$-\infty$ to $+\infty$.

Let us also assume that the central qubit system and the environment (spin chain) are initially uncorrelated with the total
density operator of the composite system at $t=-\infty$, is given by  $\rho^{tot}(-\infty)= \rho^{AB}(-\infty)\otimes \rho^E(-\infty)$, where $\rho^{AB}(-\infty)$ is the initial density operator of the two central 
qubits and  $|{\psi_E(-\infty)}\rangle$ is the initial state of the 
environment yielding the density operator $\rho^E(-\infty)= |{\psi_E(-\infty)}\rangle \langle{\psi_E(-\infty)}|$. The time evolution of the composite system  is determined by the time evolution operator $U(t)$,
as $\rho^{tot}(t)= U(t)\rho^{tot}(-\infty)U(t)^{\dagger}$. The operator $U(t)$ \cite{yuan2007}, generated by the 
Hamiltonian (\ref{product}) is
\begin{equation}
U(t) = \sum_{\mu=1}^4 |{\varphi_\mu}\rangle \langle{\varphi_\mu}|\otimes U_E^{h_\mu}(t),
\label{time}
\end{equation}
with $U_E^{h_\mu}(t)=\exp(-iH_E^{h_\mu}(t))$. Finally, the 
reduced density matrix  of the two central qubits at an instant $t$ is obtained by tracing over the environmental degrees of freedom, 
i.e., $\rho^{AB}(t)=  Tr_E[\rho^{tot}(t)]$. 
We further assume that the initial state of the two central qubits is a Werner state given by
\begin{equation}
\rho^{AB}(-\infty)= \frac{1-a}{4}I^{AB} + a |{\phi}\rangle \langle{\phi}|,
\label{initial}
\end{equation}
where $|{\phi}\rangle= (|{++}\rangle+|{--})\rangle/\sqrt{2}$. Clearly,  the werner state (\ref{initial}) becomes totally mixed for $a=0$, 
and  reduces to a pure state $|{\phi}\rangle$ in the case of $a=1$.

Noting that the coupling between the qubits and the environment provides two channels
of evolution of the environmental wave function,
the reduced density matrix of the two central qubits can now be written as \cite{liu2010}
\begin{equation}
\rho^{AB}(t) = \frac{1}{4}
\left(
\begin{array}{cccc}
1+a & 0 & 0 & 2a\sqrt{|D(t)|}\\
0 & 1-a & 0 & 0\\
0 & 0 & 1-a & 0\\
2a\sqrt{|D(t)|} & 0 & 0 & 1+a\\
\end{array}
\right),
\label{rdm1}
\end{equation}
where where the decoherence factor $D(t)= |\langle \psi_+(t)|\psi_-(t)\rangle|^2$
with
$|{\psi_+(t)}\rangle$ and $|{\psi_-(t)}\rangle$ being
 the  
states of environmental spin chain at time $t$ evolving with the Hamiltonian $H_E(h(t)+\delta)$ and
$H_E(h(t)-\delta)$, respectively \cite{damski11}.

 To evaluate $D(t)$, we rewrite the Hamiltonian (\ref{h1})
with modified $h$ (due to the coupling $\delta$)
in terms of JW fermions which then  can be decoupled into a sum of
independent $(2 \times 2)$ Hamiltonians in the Fourier space 
\cite{lieb61}.
In the basis $|{0}\rangle$ and $|{k,-k}\rangle$, which represent no quasiparticle, and  quasiparticles with momentum $k$ and $-k$,
respectively, the environmental Hamiltonian can be split as $H_E^\pm(t)=\sum_k H_k^\pm(t)$ where
\begin{eqnarray}
H_k^\pm (t)= 2 
 \left(
 \begin{array}{cc}
    h(t) \pm \delta + \cos k & \gamma \sin k  \\
\gamma \sin k & -(h(t) \pm \delta + \cos k)   \\
 \end{array}
 \right).
 \label{eq_math}
 \nonumber
\end{eqnarray}

The general wave function $\psi_{\pm}(t)$ can therefore be written as   (see Sec.{\ref{sec_txy})
\begin{equation}
|\psi_{\pm}(t)\rangle = \prod_k |\psi_k^\pm(t)\rangle 
= \prod_{k>0} \left[u_k^\pm(t) |0\rangle + v_k^\pm(t)|k,-k\rangle \right].
\end{equation}
The coefficients $u_k^\pm$ and $v_k^\pm$ are obtained by solving the Schr\"odinger
equation 
$i{\partial}/{\partial t} \left(u_k^\pm(t),v_k^\pm(t)\right)^{T}= H_k^\pm(t) \left(u_k^\pm(t),v_k^\pm(t)\right)^T$
where $A^T$ represents the transpose operation of the row matrix $A$.
Hence, the expression of $D(t)$ is given by 
$\prod_k F_k(t)=  \prod_k |\langle {\psi_k(h(t)+\delta)}|{\psi_k(h(t)-\delta)}\rangle|^2$, or,
\begin{eqnarray}
D(t)=\exp\left[
\frac{N}{2\pi} \int_0^{\pi} dk~\ln F_k
\right]
\label{eq_gendecoh}
\end{eqnarray}
where $F_k$ can be written in terms of $u_k^\pm$ and $v_k^\pm$. One finds the exact form of $F_k$ to be given by \cite{damski11,nag12}
\begin{equation}
F_k(t) \approx 1-4\sin^2(4t\delta)
\left(e^{-2\pi\tau k^2}-e^{-4\pi\tau k^2}\right).
\label{deco_momen1}
\end{equation}
In weak coupling limit ($\delta \to 0$), one can also get a closed form expression of decoherence factor (\ref{eq_gendecoh}) after crossing the first QCP at $h=1$  
given by \cite{damski11,nag12}
\begin{equation}
D(t) \approx \exp \left(-\frac{8(\sqrt2 -1)N\delta^2 t^2}{\pi \sqrt{\tau}} \right)\zeta, 
\label{weak1}
\end{equation}
where $\zeta$ is the contribution  due to the fidelity factor for adiabatically evolving modes.

\begin{figure} 
\centerline{\psfig{file=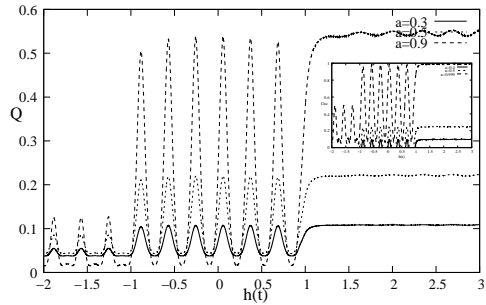,width=2.50in}}
\vspace*{8pt}
\caption{Quantum discord is plotted as a function of transverse field $h(t)$ with three different values of $a$; $a=0.9$(dashed line) corresponds to the upper one, $a=0.5$(dotted line) corresponds to the middle one, and $a=0.3$(solid line) is represented by the lower one. We have chosen $N=500$, $\delta=0.01$ and $\tau=250$. Inset shows the variation of concurrence as a function of transverse field $h(t)$.}
\label{Fig:dis_ld}
\end{figure}

 We  shall study below the behavior of discord and 
concurrence using different quenching schemes a function of $a$ and $h(t)$, as the environment is driven across QCPs. This study reveals  that the quantum discord and concurrence follow qualitatively similar kind of behavior as that of the decoherence
factor of the central qubits.
We  show that the purity of the central spin state is not entirely  determined by $a$, rather it does also depends on the decoherence factor.
The central qubits will be in a pure state only when $a=1$ as well as $D(t)=1$.

\begin{figure} 
\centerline{\psfig{file=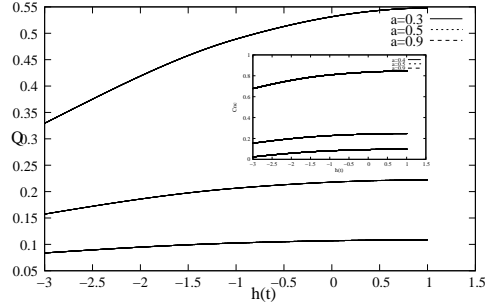,width=2.50in}}
\vspace*{8pt}
\caption{Quantum discord is plotted as a function of transverse field $h(t)$. $a=0.9$(dashed line) corresponds to the upper one, $a=0.5$(dotted line) corresponds to the middle one, and $a=0.3$(solid line) is represented by the lower one. We take $N=500$, $\delta=0.0001$ and $\tau=250$. Inset shows the variation of concurrence as a function of transverse field $h(t)$.}
\label{Fig:dis_sd}
\end{figure}

\begin{figure} 
\centerline{\psfig{file=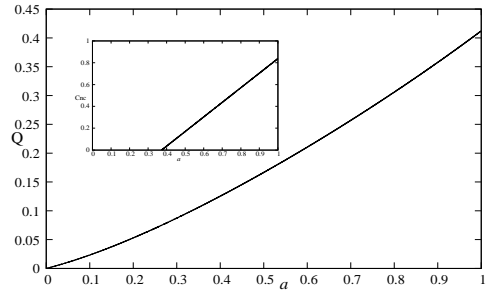,width=2.50in,height=1.50in}}
\vspace*{8pt}
\caption{Concurrence and discord are plotted as a function of $a$. Here decoherence factor $D(t)=0.7025$ was taken from transverse Ising quenching data, $\tau=250$, $\delta=0.0001$ and $t=251$,  i.e., after crossing the first Ising quantum critical point at $h(t)=1$. }
\label{Fig:comp}
\end{figure}
 
Using the reduced matrix Eq.~(\ref{rdm1}), one can find the quantum discord in the form \cite{liu2010}, 
 \begin{equation}
Q[\rho^{AB}(t)]= - \frac{1+a}{2}\log_{2}(\frac{1+a}{4})+ \sum_m^4 \lambda_m \log_2 \lambda_m + f(a),
\label{dis}
\end{equation}
where $\lambda_m$'s are the eigen value of $\rho^{AB}(t)$ and $f(a)$ is a function of $a$.
Similarly, the concurrence\cite{liu2010}
can be obtained as
\begin{equation}
C_{nc}[\rho^{AB}(t)]= max[a(\sqrt{|D(t)|}+ \frac{1}{2})-\frac{1}{2},0].
\label{cnc}
\end{equation}

The equilibrium behavior of discord between two central qubits coupled to a transverse spin chain has been extensively studied \cite{liu2010}. We would like to investigate large $\delta$ behavior of discord and concurrence  numerically studying Eqs.~(\ref{dis}) and (\ref{cnc})
for transverse quenching across Ising critical points achieved by varying $h=1- t/\tau$. Fig.~(\ref{Fig:dis_ld}) shows that discord profile 
has repeated collapses and revivals similar to that of decoherence factor as described in the paper by  Damski $\it et~al$ \cite{damski11}.
We see complete revival in the region bounded by the Ising critical points $h=\pm 1$, while beyond the critical point at $h=-1$, one observes
 partial revivals.
It is important to note that the  discord  between the two central spins does never vanish even if we start with an initial state with   $a \to 0$, when
the qubits are in a complete mixed state.
The qualitative profile of concurrence as a function of $h(t)$ is shown in the inset; this turns out to be similar to that of discord.  
The quantitative difference when  compared to the behavior of discord is the followoing: 
concurrence has a higher magnitude compared to that of discord for the same value of $a$ and secondly, the concurrence between two spins vanishes completely 
after crossing the second QCP after certain value of $a$ which is unlikely as compared to the discord.  Concurrence stays zero throughout the quenching time below a 
certain value of $a$ while discord is always non-zero. This is identical to the situation when the concurrence and discord are studied
in the final state following a quantum quench discussed in previous sections.

Let us now proceed to the small $\delta$ limit when one finds that both  discord and concurrence exhibit  decay with time
after crossing QCP $h=1$ which is again  similar to the behavior of the  decoherence factor in the same limit \cite{damski11} (see Fig.~(\ref{Fig:dis_sd})).
 The quantitative
difference between two measures can be understood using similar arguments as given above.  For example,
in Fig.~(\ref{Fig:comp}), we compare the behavior of discord and concurrence as a function of $a$ after 
crossing the first QCP, i.e., for $h(t)<1$. We find that the concurrence becomes linear in $a$ beyond a threshold  
value of $a$ when the central spins have a non-zero concurrence; discord is always non-zero for all 
values of $a$. These results bear close similarity with the corresponding equilibrium study
\cite{liu2010}. Identical results are obtained when the environment is quenched across
anisotropic critical points (ACPs) and multicritical points (MCPs) using appropriate quenching schemes and interaction Hamiltonians between the qubits and the spin chain.

\section{Conclusions}
This review attempts to provide a bridge between non-equilibrium dynamics of quantum critical
systems and quantum information theory. We have shown that quenching through QCPs leads
to defects in the final states; these defects in fact lead to quantum correlations which are manifested
in measures like concurrence and discord. At the same time both these measures appear
to satisfy the same universal scaling relation as that of the defect density. Moreover, under certain
conditions the final state has zero concurrence but non-zero discord implying that correlation generated through quench  is entirely quantum mechanical. We believe these observations will lead to future studies involving dynamics and quantum information.

Let us  conclude with some generic comments: models discussed in the review are integrable and also reducible to decoupled two-level systems for which the non-adiabatic transition probabilities can be analytically obtained using the Landau-Zener transition formula. Moreover,
the reduced matrix has a simplified structure due to different symmetry properties of the underlying
Hamiltonian. Question therefore remains what would happen
when one studies quenching of a non-integrable model and probe correlations in the final state. Do concurrence or discord present in the final state satisfy a universal scaling relation? If so, is the scaling identical to that of defect density? How do they evolve   when the system thermalizes
following a quench? These open  questions are to  be addressed in future studies in this fertile
area of research.

\section*{Acknowledgements}

We are very much thankful to Dr. Uma Divakaran and Shraddha Sharma for useful comments. AD
acknowledges CSIR, New Delhi, for financial support.

\end{document}